\documentstyle[preprint,psfig,aps]{revtex}
\begin{document}                
\def\a{$\alpha$}
\def\be{\begin{equation}}
\def\ee{\end{equation}}
\def\ba{\begin{eqnarray}}
\def\ea{\end{eqnarray}}
\preprint{DOE/ER/40561-58-INT99}
\title{A classical two-body Hamiltonian model and its mean field approximation}
\author{George F. Bertsch, Thomas Papenbrock, and Sanjay Reddy}
\address{Institute for Nuclear Theory, Department of Physics, 
University of Washington, Seattle, WA 98195, USA}
\maketitle
\begin{abstract}
We extend a recent billiard model of the nuclear $N$-body Hamiltonian to 
consider a finite two-body interaction.  This permits a treatment of
the Hamiltonian by a mean field theory, and also allows the possibility
to model reactions between nuclei. The density and the mean field potential 
can be 
accurately described by a scaling function which shows the qualitative 
features of the liquid drop picture of the nucleus. 
\end{abstract}
The nuclear many-body problem is often approximated by phenomenological mean
field or liquid drop models which serve as good starting points to study the
shape and the spectrum of low lying single particle and collective excitations
\cite{BohrM,RingS,BertschB}. This well-known approach is a consequence of the
inadequate knowledge of the microscopic nuclear Hamiltonian, and the difficulty
in dealing with the many-body problem. The mean field approximation is commonly
employed to make the many-body problem tractable.  Therefore it is of interest
to question of how well mean-field theory works for a given microscopic
Hamiltonian that can be solved exactly.

A promising model for such a program may readily be obtained from a recent
billiard model of the nuclear $N$-body Hamiltonian. This is an interacting
$N$-body system with two-body interactions that is rich enough to show various
features of self-bound many-body systems yet simply enough to allow for
practical calculations and an understanding \cite{TP}.

Generalizing the model of ref. \cite{TP}, we consider the Hamiltonian
\be
\label{ham}
H=\sum_{i=1}^N{p_i^2\over 2m} + \sum_{i<j}V(|\vec{r}_i-\vec{r}_j|),
\ee
where $\vec{r}_i$ is a three-dimensional position vector of the $i$-th nucleon
and $\vec{p}_i$ is its conjugate momentum and the interaction given by
\ba
\label{int}
V(r)=\left\{
     \begin{array}{ll}
     -V_0 & \mbox{for $r<a$}, \\
     0 & \mbox{for $r\ge a$},
     \end{array}
     \right.
\ea
where $V_0$ is a positive constant. Total energy $E$, momentum $\vec{P}$ and
angular momentum $\vec{L}$ are conserved quantities. Interactions alter
the momenta of particles whenever they are 
distance $a$ apart, and otherwise they move freely.

For $V_0=\infty$ one obtains a billiard model, which was recently studied in
ref. \cite{TP,TPTP}.  Due to the simple form of the two-body interaction, the
numerical integration of the equations of motion is quite easy, scaling with
particle number as $N \log N$ rather than the usual $N^2$ typical of the
classical $N$-body problem.  It was found that the dynamics is dominantly
chaotic and ergodic. In the center of mass system one finds a constant
single-particle density inside a circle of diameter $a$ that quickly drops to
zero within a thin surface region. The phase space structure does not depend on
energy in such billiards.

Finite values of $V_0$ introduce an energy scale and a finite binding energy of
the system.  The Hamiltonian then permits reactions like particle capture or
emission, giving it important features for a model of the nucleus and its
dynamics.
 
In this note we will examine the mean field theory of of the $N$-body
Hamiltonian~(\ref{ham},\ref{int}). To this purpose we introduce the
single-particle phase space density $f(\vec{r},\vec{p})$.  Let
$p^2/2m+W(\vec{r})$ be the mean field Hamiltonian.  As is well known, any
density which is a functional of the mean field Hamiltonian solves the
corresponding Vlasov equation \cite{Liboff}.  However, we are interested in the
situation corresponding to thermal equilibrium and choose the canonical
distribution function \be
f(\vec{r},\vec{p})\propto\exp{\left(-{p^2/2m+W(\vec{r})\over kT}\right)}, \ee
where $T$ denotes the temperature.  Obviously, the phase space distribution
function is a product of the normalized momentum space density $\left[4(2\pi
  mkT)^{3/2}\right]^{-1} \exp{\left(-p^2/2mkT\right)}$ and the number density
\be
\label{ndens}
n(\vec{r})=c\,\exp{\left(-{W(\vec{r})\over kT}\right)}.
\ee
The normalization $c$ is given by
\[ c=N\left[\int d^3r\,n(\vec{r})\right]^{-1} \] 
The mean field potential $W(\vec{r})$ depends on the number density and
the two-body interaction
\be
\label{vmean}
W(\vec{r})=\int d^3x\,V(|\vec{r}-\vec{x}|)\,n(\vec{x}).
\ee
Eq.(\ref{ndens}) together with eq.(\ref{vmean}) are the mean field
equations to be solved self-consistently.

We next observe that the solution of the mean field equations 
(\ref{ndens}-\ref{vmean}) 
will have an interesting scaling property.
Let $W(\vec{r})$ and $n(\vec{r})$ denote the mean field potential and 
density for parameters 
$N,V_0$ and $kT$. Then $\alpha W(\vec{r})$ is the mean field potential for
parameters $\alpha NV_0$ and $\alpha kT$, while $\alpha n(\vec{r})$ is the
density for parameters $\alpha N$ and $V_0/\alpha kT$. 
Thus, the normalized quantities $V(\vec{r})/NV_0$ and $n(\vec{r})/N$ depend 
only on the ratio 
\be
\gamma\equiv NV_0/kT
\ee 
of parameters. This scaling behavior comes directly from eqs.
(\ref{ndens},\ref{vmean}) and does not depend on the 
specific form of the two-body interaction (\ref{int}).

It is well known that the density of an open system can be represented as a 
thermal distribution function only for $\gamma \gg 1$. In this case the 
density is exponentially small outside the nucleus and can be omitted. In what
follows we restrict ourselves to this low temperature regime. 
  
We are interested in spherical symmetric solutions of the
mean field equations. In this case eq.(\ref{vmean}) may be written as
\be
W(r)=\int\limits_0^\infty dx\, n(x) v(r,x),
\ee
where
\ba
v(r,x)=-2\pi V_0\left\{
     \begin{array}{ll}
     0 & \mbox{for $a<|r-x|$}, \\
     2x^2 & \mbox{for $a>r+x$},\\
     {x\over 2r} \left[a^2-(r-x)^2\right] & \mbox{for $|r-x|<a<r+x$}.
     \end{array}
     \right.
\ea
For computational purposes it is useful to rewrite the mean field potential as
\ba
\label{compot}
W(r)=-2\pi V_0\Bigg[&& \Theta(a-r)\int\limits_0^{a-r}dx\,2x^2\,n(x) \nonumber\\
&+& 
\Theta(a-r)\int\limits_{{\rm max}(a-r,r)}^{a+r} dx\,{x\over 2r}(a^2-(x-r)^2)
\,n(x)\nonumber\\
&+&\Theta(r-a/2)\int\limits_{|r-a|}^r dx\,{x\over 2r}(a^2-(x-r)^2)\,n(x)\Bigg]
\ea
where $\Theta(x)$ is the unit step function.

\begin{figure}
  \begin{center}
    \leavevmode
    \parbox{0.9\textwidth}
           {\psfig{file=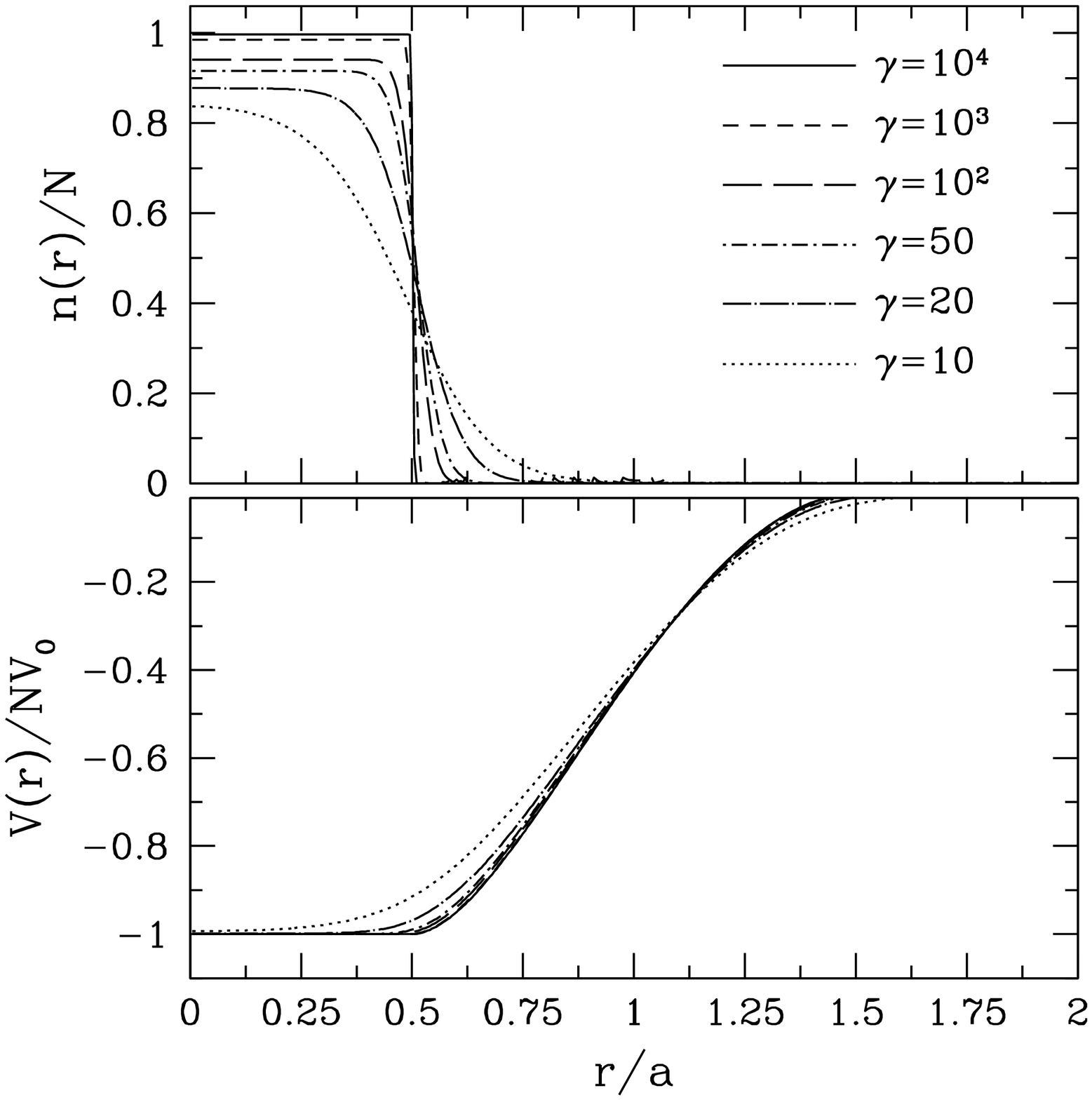,width=0.85\textwidth,angle=0}}
  \end{center}
\protect\caption{Mean field potentials (lower panels) and densities 
(upper panels) for various values of the parameter $\gamma=NV_0/kT$. }  
\label{fig1}
\end{figure}

We solve the mean field equations numerically by iteration. We take an
initial density $n(r)$ that is constant inside a sphere of diameter $a$
and zero outside. Upon several iterations one obtains a converged solution.
Fig.~\ref{fig1} shows the resulting number density $n(r)/N$ and mean field
potential $W(r)/NV_0$ for different values of the parameter 
$\gamma$. For large $\gamma$ the mean field potential is flat 
for $r<a/2$ and increases until $r\approx 3a/2$ while the density is constant
for $r< a$ and quickly drops to zero around $r\approx a/2$. 
Decreasing values of $\gamma$ (i.e. decreasing values of $V_0$ and/or 
increasing temperatures $T$ while the number of particles $N$ is kept fixed)
lead to a thicker surface region and to a more extended mean field potential. 
However, for $r<a$ the potential is still a flat bottom potential and the
density is roughly constant. Thus, the system exhibits qualitative features 
of the liquid drop model. However, our simple model does not yield 
a saturation of the density or the binding energy with increasing $N$. 
This is due to the absence of a repulsive potential core.

In the limit $\gamma \to \infty$ the density approaches that of
a sharp-edged liquid drop, constant inside a sphere of radius
$a/2$ and zero outside.  For finite values of $\gamma$, the 
surface thickness can be estimated assuming a trapezoidal density
distribution.  Taking the width parameter as $\epsilon$ and demanding
self-consistency in the neighborhood of $r=a/2$, one finds the relation
\be
\varepsilon/a=(3\gamma)^{-1/2}.  
\label{eps/a}
\ee

\begin{figure}
  \begin{center}
    \leavevmode
    \parbox{0.9\textwidth}
           {\psfig{file=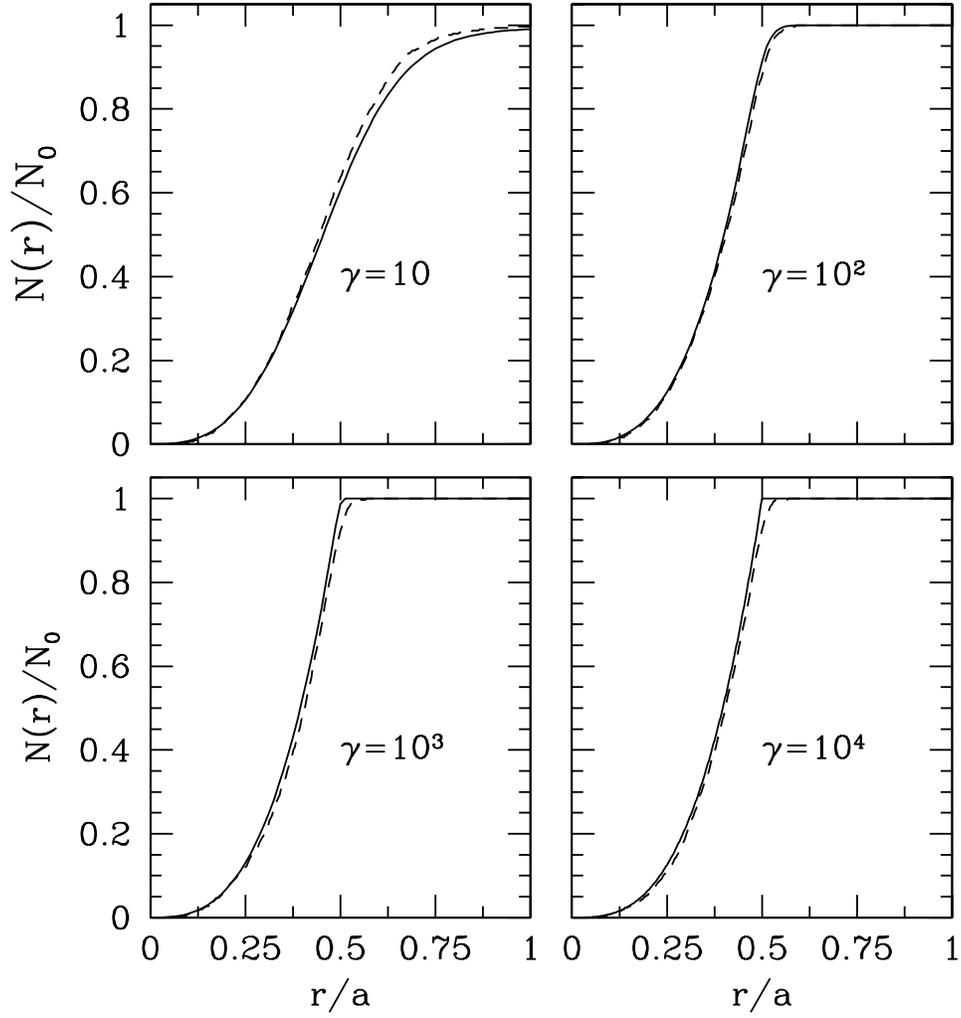,width=0.85\textwidth,angle=0}}
  \end{center}
\protect\caption{Integrated number density from microscopic calculations
(dashed lines) compared to mean field results (solid lines) for different 
values of the parameter $\gamma=NV_0/kT$}  
\label{fig2}
\end{figure}

It is also interesting to compare the mean field results with the interacting
system (\ref{ham},\ref{int}). Agreement may only be expected for sufficiently
large numbers of particles $N_0$. In what follows we set $N_0=300$.
The initial conditions of the $N_0$-body system are chosen such that the 
positions are drawn at random inside a sphere of diameter $a$ while 
the momenta are drawn from a Maxwell-Boltzmann distribution with 
temperature $T$. We follow the time evolution of the many-body system for
roughly $10^5$ interactions to allow for equilibration. At the end
of the evolution we take the positions in the center of mass system
and compute the integrated density
\be
N(r)=4\pi\int\limits_0^r dx\,x^2 n(x)
\ee
by counting the number of particles inside a sphere of diameter $r$.
Fig.~\ref{fig2} compares the average of ten runs with the mean field
result for various values of the temperature. The agreement is rather
good confirming the validity of the mean field approximation.

Finally, we ask what values the parameters should have to correspond
to physical properties of nuclei.  Let us consider a nucleus of mass
$A\approx 75$.  Then the radius is given by $a/2\approx 1.2 A^{1/3}
\approx 5$ fm.  The liquid-drop surface thickness is roughly
$\epsilon \approx 1$ fm, yielding from eq. (\ref{eps/a}) $\gamma\approx
30$.  We would like to set the energy scale by the depth of the
mean field potential, which should be $N V_0 \approx 50$ MeV to 
correspond to a typical Woods-Saxon potential of nuclear physics.
Turning to eq. (7), we see that the two parameter values can be
obtained taking the temperature as $kT\approx 2$ MeV.  Equilibrated
nuclei at such a temperature can be easily produced in heavy ion
reactions, so the model might have some applicability to nuclear
reactions.  However, in detail the Fermionic nature of the nuclear
many-body problem will make considerable differences from the
present classical model.  In particular, the chemical potentials
at a given temperature are very different, which would be important
for the nucleon evaporation rates.
\section*{Acknowledgments}
This work was supported by the Department of Energy under grant 
DOE/ER/40561.

\end{document}